\documentclass[prd,twocolumn,nofootinbib]{revtex4}
\usepackage{graphicx, epsfig}
\usepackage{color}
\usepackage{mathrsfs}
\usepackage{bm}
\usepackage{amsmath,amssymb,empheq}
\usepackage{mathrsfs}
\usepackage[caption=false]{subfig}
\usepackage{hyperref}
\usepackage[normalem]{ulem}
\usepackage{mathtools}
\usepackage[x11names]{xcolor}

\definecolor{fashionfuchsia}{rgb}{0.96, 0.0, 0.63}
\colorlet{no_so_fashion_purple}{blue!50!red}

\newcommand{\be}{\begin{equation}}
\newcommand{\ee}{\end{equation}}
\newcommand{\ba}{\begin{eqnarray}}
\newcommand{\ea}{\end{eqnarray}}

\newcommand{\nn}{\nonumber}

\newcommand{\la}{\langle}
\newcommand{\ra}{\rangle}

\begin{document}
\title{Time-Dependent Tunneling in the Thin-Barrier Limit}
\author{Tanmay Vachaspati$^a$\footnote{tvachasp@asu.edu}, Frank Wilczek$^{a,b,c,d,e,f}$\footnote{fwilczek@asu.edu},
Zara Yu$^f$\footnote{zarayu@mit.edu}}
\affiliation{
$^a$Department of Physics, Arizona State University, Tempe, Arizona 85287, USA}
\affiliation{$^b$T. D. Lee Institute, Shanghai 201210, China}
\affiliation{$^c$Wilczek Quantum Center, Department of Physics and Astronomy, Shanghai Jiao Tong University, Shanghai 200240, China}
\affiliation{$^d$Department of Physics, Stockholm University, AlbaNova University Center, 106 91 Stockholm, Sweden}
\affiliation{$^e$Nordita, Stockholm University and KTH Royal Institute of Technology, Hannes Alfv\'{e}ns v\"{a}g 12, SE-106 91 Stockholm, Sweden}
\affiliation{$^f$Center for Theoretical Physics, Massachusetts Institute of Technology, Cambridge, Massachusetts 02139, USA}

\begin{abstract}
The usual WKB analysis for quantum tunneling applies when the tunneling action is large, as it is for tall, wide potential barriers. In contrast we analyze tunneling when the action is small, as it is for tunneling across a tall, thin barrier. We develop a perturbative analysis where the control parameter is the inverse of the area under the potential barrier and apply our technique to several examples in $1+1$ dimensions. In resonant situations for bound particles we find that the tunneling probability grows with time as $\propto t^2$, while in non-resonant situations it grows linearly with time. We evaluate not only the tunneling probability but also the time-dependent tunneling  wavefunction for a particle that escapes to infinity, {\it i.e.} from a quasi-bound state to  the continuum.
\end{abstract}

\maketitle

Quantum mechanical tunneling is usually discussed as a time-independent problem using the WKB approximation, simply in terms of the probability of transmission through a barrier~\cite{Messiah_Albert, Griffiths_Schroeter_2018}. While this approach has widespread application, it loses dynamical information about the tunneling and its dependence on the initial state of the bound particle. Further, the WKB approximation is controlled when the action under the barrier is large, as commonly occurs for sufficiently tall, thick barriers. In some applications this condition is not met and a different approach is necessary.

Here we analyze tunneling through a tall, thin barrier. Though the WKB approximation does not apply, the situation still involves quantum tunneling, since a classical particle would not be able to pass through the tall barrier. Earlier work considering the tall, thin barrier has focused on the transmission of incoming plane waves through the barrier~\cite{Elberfeld_Kleber_1988}. The solutions can be expressed in terms of Moshinsky functions, which were originally introduced in the context of diffraction in time~\cite{Moshinsky_1952} and have since been used in the broader study of time-dependent quantum phenomena~\cite{Kleber_1994, Del_Campo_2009}. Our analysis focuses on the decay of bound states and uses a perturbative approach in the strength of the tall, thin barrier. We work through several examples and obtain the full time dependent tunneling wavefunction and the time dependent tunneling probability. We do not claim that delta-function barriers or time-dependent tunneling transients are new. Rather, we use the tall/thin barrier limit as a simple asymptotic organization of several tunneling problems that are often discussed separately: resonant coherent transfer, nonresonant leakage, and escape into a continuum.

In Sec.~\ref{tallthin} we discuss the delta function barrier as a limiting case of a tall, thin rectangular barrier. This exercise determines the coefficient of the delta function in the Hamiltonian in terms of the height and width of the potential barrier. In Sec.~\ref{general} we consider symmetric potential wells, $V(x)=V(-x)$, with the delta function barrier at $x=0$. The symmetry allows for wavefunctions to be classified by their parity. The general results developed in Sec.~\ref{general} are used to solve for tunneling for a square well and a quadratic potential in Sec.~\ref{examplessymmetric}. In Sec.~\ref{asymV} we discuss tunneling in asymmetric situations involving a square well potential. If the delta function divides the square well into two wells, one of width $a$ and the other of width $b$, important aspects of the tunneling depend on whether $b/a$ is rational or irrational, since this determines whether energy levels in one well match energy levels of the second well. Since we have already dealt with the symmetric case when $b=a$ in Sec.~\ref{general}, we focus on the case when $b/a$ is irrational. In Sec.~\ref{escape} we consider a particle trapped between two equal delta function barriers that eventually tunnels and escapes to infinity. We evaluate both the tunneling rate and the tunneling wavefunction in the approximation that the tunneling is suppressed. In Sec.~\ref{other_approaches} we discuss connections with other approaches to time-dependent tunneling. We summarize our findings in Sec.~\ref{conclusions}.

As later will be elaborated in context, the formulae below should be understood within their corresponding time windows. The quadratic behavior in the resonant double-well problem is the early-time limit of coherent oscillation. The linear behavior in the nonresonant case is an intermediate regime after the initial quadratic transient has been resolved but before recurrences or reflections become important. The exponential decay in the escape problem is likewise an intermediate-time approximation dominated by the metastable pole; very early and very late times generally contain non-exponential corrections~\cite{Winter_1961}.

\section{Tall, thin barrier and delta function barrier} \label{tallthin}

\begin{figure}
\includegraphics[width=0.30\textwidth,angle=0]{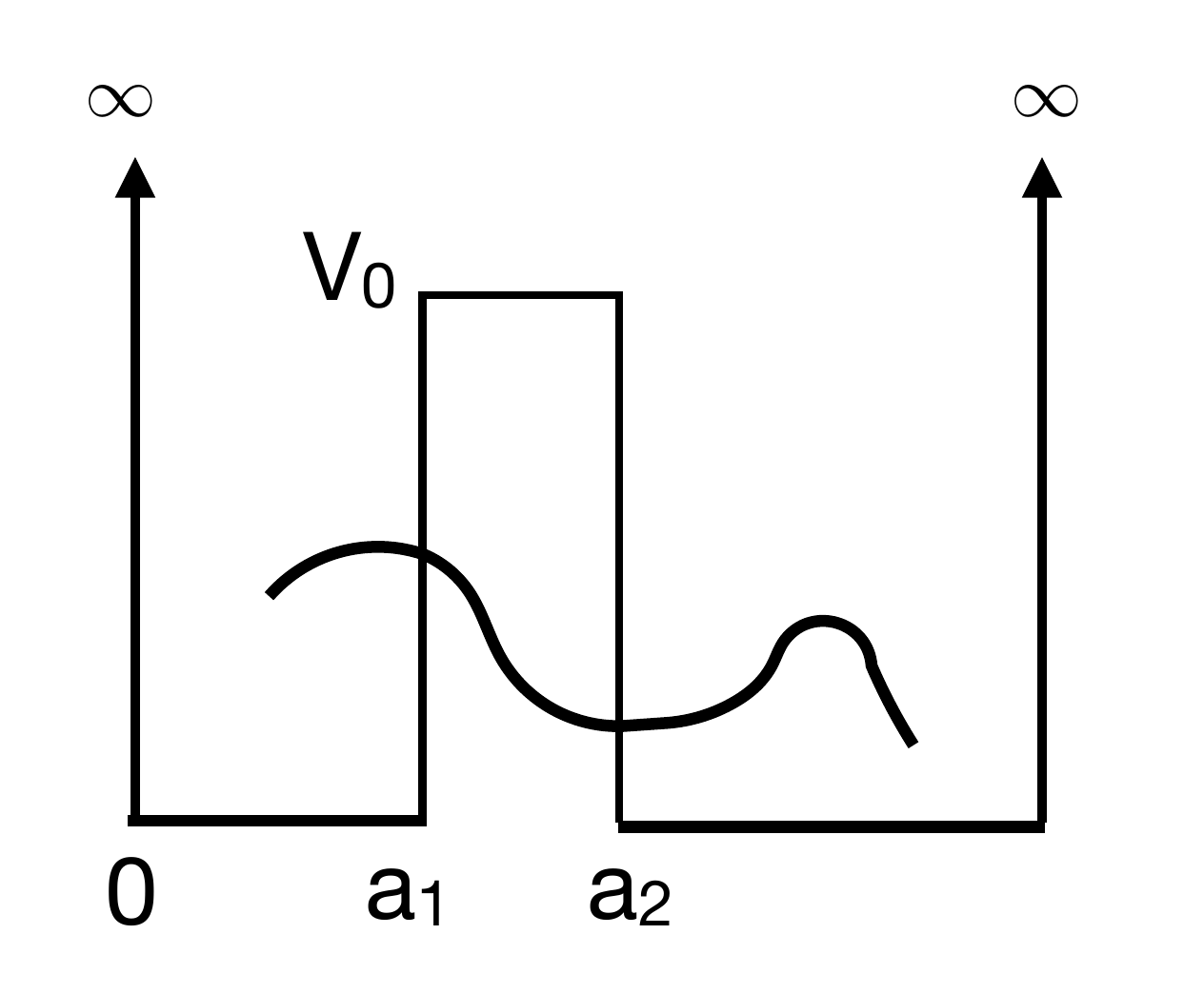}
\caption{A finite barrier of height $V_0$ and width $d=a_2-a_1$ and an illustration of part of a wavefunction. 
}
\label{finitebarrierfig}
\end{figure}

Consider the tall, thin square barrier depicted in Fig.~\ref{finitebarrierfig}. The wavefunction under the barrier for an energy eigenstate with energy $E$ is given in terms of the wavefunction and its derivative at $x=a_1$, denoted $\psi_1$ and $\psi'_1$, by
\ba 
&& \hskip -0.75 cm
\psi(x) = \psi_1 \cosh (\rho (x-a_1)) + \frac{\psi'_1}{\rho}  \sinh(\rho (x-a_1)), 
\\
&& \hskip -0.75 cm
\psi'(x) = \rho \psi_1 \sinh (\rho (x-a_1))  + \psi'_1 \cosh (\rho (x-a_1)),
\ea
where $\rho = \sqrt{2\mu (V_0-E)}$ and $\mu$ is the mass of the particle. This gives the wavefunction and the derivatives on the other side of the barrier, at $x=a_2$,
\ba
&& \hskip -0.75 cm
\psi_2 = \psi_1 \cosh (\rho d) + \frac{\psi'_1}{\rho}  \sinh(\rho d), 
\label{psi2}
\\
&& \hskip -0.75 cm
\psi'_2 = \rho \psi_1 \sinh (\rho d)  + \psi'_1 \cosh (\rho d),
\label{psi2prime}
\ea
where $d=a_2-a_1$.

We are interested in a tall, thin barrier, so $V_0 \to \infty$, $d \to 0$ with $(V_0-E) d \equiv \alpha^{-1}$ held fixed. These conditions imply
\be
\rho d = \sqrt{2\mu (V_0-E)} \, d \approx \sqrt{\frac{2\mu d}{\alpha}} \to 0. 
\label{kappad}
\ee
Therefore, from \eqref{psi2} and \eqref{psi2prime},
\ba
\psi_2 &\to& \psi_1, \\
\psi'_2 &\to& \psi'_1+ (\rho^2 d ) \psi_1 \nn \\
        &\sim& \psi'_1 + 2\mu (V_0-E) d \,  \psi_1
            = \psi'_1 + \frac{2\mu}{\alpha} \psi_1.
\ea
The latter relation can be written as
\be
-\frac{1}{2\mu} (\psi'_2-\psi'_1) + \alpha^{-1} \psi_1 =0
\ee
which is interpreted as the Schrodinger equation,
\be
{\hat H} \psi = \left [ \frac{p^2}{2\mu}  + V(x) + \alpha^{-1} \delta (x) \right ] \psi
= i \partial_t \psi,
\label{schrodinger}
\ee
integrated across the delta function barrier. Hence the limit of a very tall, thin barrier is simply that of a delta function barrier with ``strength'' given by $\alpha^{-1}=(V_0-E) d$.  Thus $\alpha$ is not the barrier area, but its inverse. The limit $\alpha \rightarrow 0$ is the opaque thin-barrier limit, and the expansion below is an expansion in the transparency of that barrier.

While the above discussion was for a tall, thin rectangular barrier, for a potential barrier of general shape we can expect
\be
\alpha^{-1} \approx \int_{a_1}^{a_2} dx\, (V(x)-E)
\label{alpha-1},
\ee
where $a_1$ and $a_2$ are turning points.

Note that the tunneling rate calculated in the WKB approximation is suppressed by $\exp(-2S)$ where the action $S$ is given by
\be
S = \int_{a_1}^{a_2} dx \, \sqrt{2\mu (V(x)-E)},
\ee
which for the rectangular barrier is $S = \rho d \approx \sqrt{2\mu (V_0-E)}\, d$. The usual WKB formula is controlled in the large-action regime, {\it i.e.} $S \gg 1$. In contrast, for the tall, thin barrier we are considering, we have $S \ll 1$ as seen in \eqref{kappad}. 

The instanton method provides another semiclassical description of tunneling, based on saddle points of the Euclidean path integral, and is similarly controlled when the relevant action is large~\cite{Coleman_1977, Callan_Coleman_1977}. In the small-action regime considered here, fluctuations are not parametrically suppressed and the semiclassical instanton expansion is therefore not controlled.

Our strategy for solving the time-dependent tunneling problem is to first solve the Schrodinger equation for $\alpha=0$ when there is no tunneling, and then to use a perturbative expansion in $\alpha$ to obtain the tunneling wavefunction from which we can calculate physical quantities of interest. We develop the general formalism in Sec.~\ref{general} for a potential that is symmetric, $V(x)=V(-x)$ (see Fig.~\ref{setupfig1}), and apply the formalism to two examples -- a square well and a quadratic potential -- in Sec.~\ref{examplessymmetric}.

\section{General formalism for symmetric potentials} \label{general}

The Hamiltonian for our setup is,
\be
{\hat H} = \frac{p^2}{2\mu}  + V(x) + \alpha^{-1} \delta (x)
\label{hamiltonian1}
\ee
where $V(x)=V(-x)$ is some symmetric potential function. For illustration purposes we will take $V(x)=x^2/2$ but our analysis applies more generally.

\begin{figure}
\includegraphics[width=0.30\textwidth,angle=0]{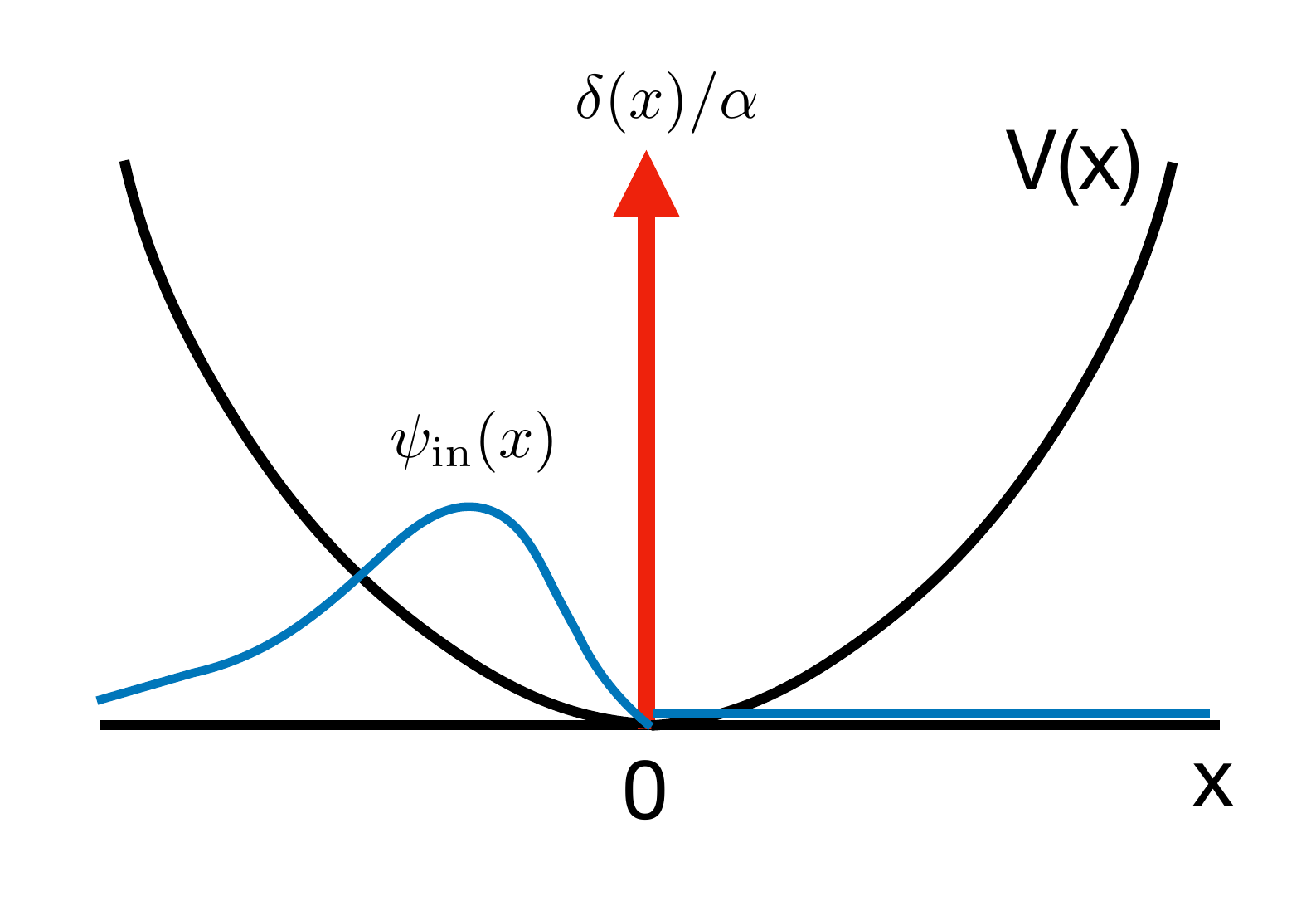}
\caption{The quadratic potential with a delta function barrier in the middle. The wave function is initially localized to the left of the barrier but tunnels to the right of the barrier as time goes on.
}
\label{setupfig1}
\end{figure}

We will solve the time-dependent Schrodinger equation 
\be
{\hat H} \Psi = i \partial_t \Psi
\label{schrodinger1}
\ee
with initial conditions so that the particle is localized to the left side of the delta function barrier (see Fig.~\ref{setupfig1}).

\subsection{Initial state} \label{initial}

We would like to start off with a particle located in the left well and calculate the probability of finding it in the right well. The initial wavefunction, $\psi_{\rm in}(x)$, is illustrated in Fig.~\ref{setupfig1}. We take it to be in the ground state of the Hamiltonian in \eqref{hamiltonian1} but with $\alpha=0$, with the wavefunction localized entirely in the $x < 0$ region. In other words, $\psi_{\rm in}$ satisfies
\be
- \frac{1}{2\mu}\partial_x^2  \psi_{\rm in}  + V(x)  \psi_{\rm in} = E_{\rm in} \psi_{\rm in} ,
\ \ x < 0
\ee
with the boundary conditions
\be
\psi_{\rm in} (-\infty)=0, \ \  \psi_{\rm in}(0) = 0
\ee
and $\psi(x)=0$ for $x \ge 0$. The energy eigenvalue $E_{\rm in}$ is left unspecified for now. The initial wavefunction is assumed to be properly normalized,
\be
\int dx \,  |\psi_{\rm in}|^2 =1.
\ee

\subsection{Eigenstates} \label{eigenstates}

\begin{figure}
\includegraphics[width=0.30\textwidth,angle=0]{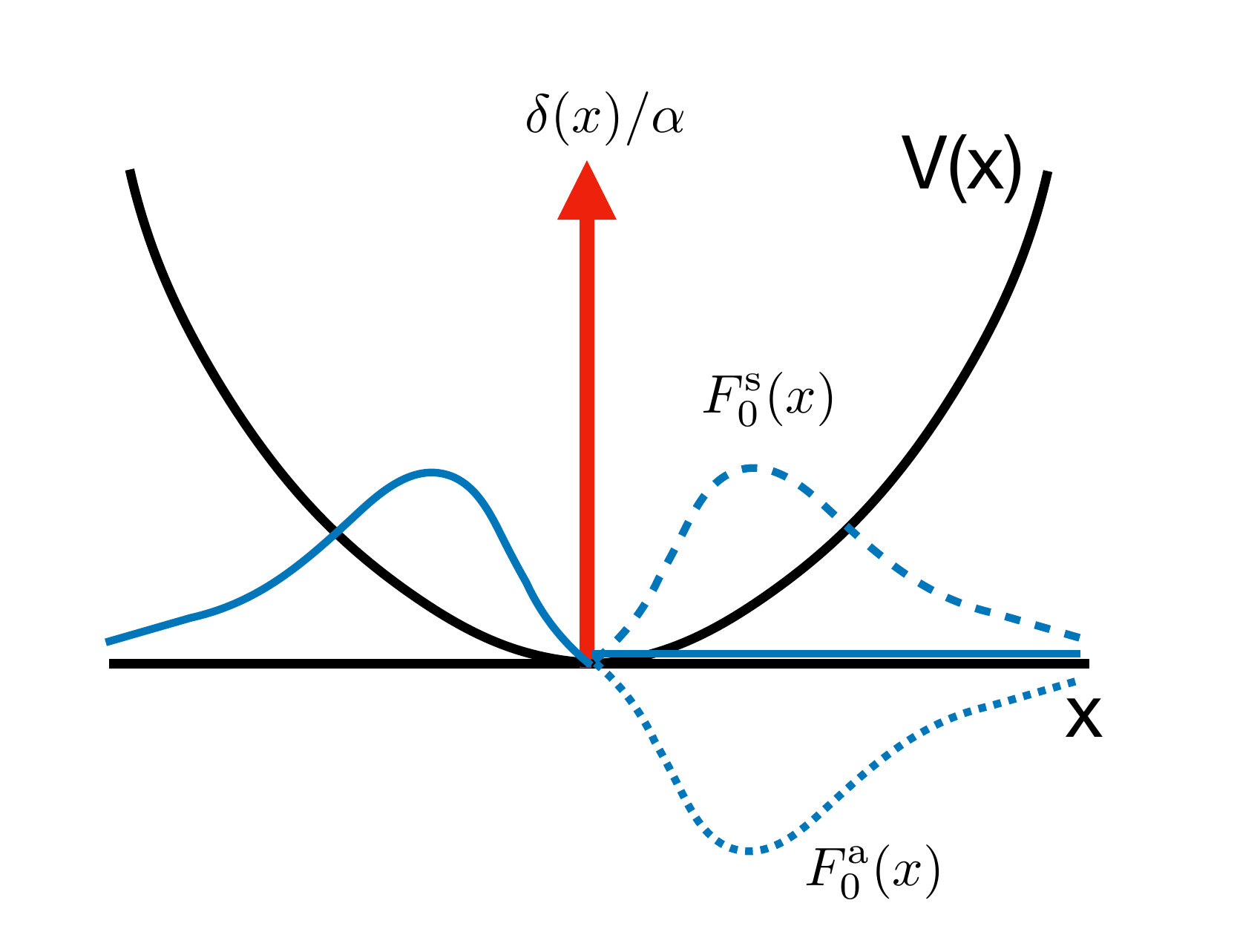}
\caption{Lowest energy wavefunctions for $\alpha =0$.
}
\label{setupfig2}
\end{figure}

The eigenstates with $\alpha=0$ can be written in terms of symmetric and antisymmetric combinations of the wavefunctions in the left- and right- sides of the potential. We denote the energy eigenstates for $\alpha=0$ by $F_n^{\rm (s)}(x)$ and $F_n^{\rm (a)}(x)$,
\ba
F_n^{\rm (s)}(x) &=& \frac{1}{\sqrt{2}} [ f_n (x) + f_n(-x) ], \label{Fns} \\
F_n^{\rm (a)}(x) &=& \frac{1}{\sqrt{2}} [ f_n (x) - f_n(-x) ],
\ea
where $f_n(x)$ is the normalized $n^{\rm th}$ energy eigenstate for $x < 0$ and $f_n(x)=0$ for $x \ge 0$. Note that $\psi_{\rm in}(x) = f_0(x)$. The functions $F_n^{\rm (s)}(x)$ and $F_n^{\rm (a)}(x)$ are illustrated in Fig.~\ref{setupfig2}.

\begin{figure}
\includegraphics[width=0.30\textwidth,angle=0]{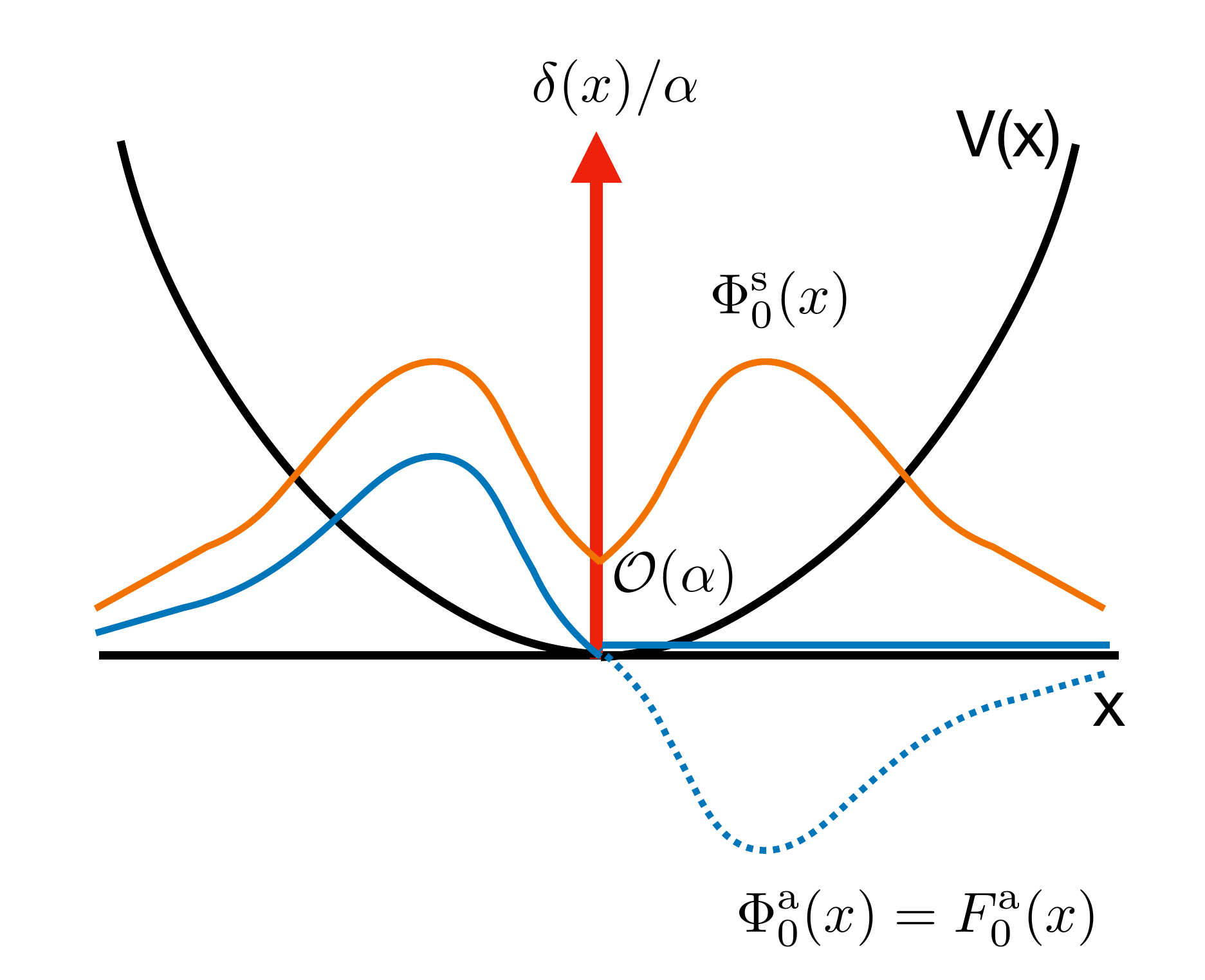}
\caption{Lowest energy wavefunctions for small but non-zero $\alpha$.
}
\label{setupfig3}
\end{figure}

Next we consider the symmetric and antisymmetric eigenstates, $\Phi_n^{\rm (s)}(x)$ and $\Phi_n^{\rm (a)}(x)$, for $\alpha \ne 0$ and $\alpha \ll 1$. The effect of increasing $\alpha$ is to change the boundary conditions for the symmetric $F_n^{\rm (s)}(x)$ because integration
of \eqref{schrodinger} across the delta function barrier gives,
\be
\psi (0) = \frac{\alpha}{2\mu} [ \psi'(0+)-\psi'(0-) ]. 
\label{psi0}
\ee
Hence the eigenstates for non-zero $\alpha$ are (see Fig.~\ref{setupfig3}),
\ba
\Phi_n^{\rm (s)} (x) &=& F_n^{\rm (s)}(x) + \alpha g_n(x), \label{Phins} \\
\Phi_n^{\rm (a)} (x) &=& F_n^{\rm (a)}(x) \label{Phias},
\ea
where $g_n(x)$ can be found by solving the Schrodinger equation, taking into account the boundary condition in \eqref{psi0}. However, we leave $g_n$ as an unspecified function and return to it at a later point. The antisymmetric states are unaffected by the presence of the delta function barrier.

Now the initial quantum state can be expanded in the $\Phi_n^{\rm s,a}$ basis,
\be
|\psi_{\rm in} \ra = \sum_{n,i} |\Phi_n^{(i)}\ra \la \Phi_n^{(i)} | \psi_{\rm in} \ra
\ee
where $i = s, a$. Schrodinger evolution gives
\be
|\psi (t) \ra = \sum_{n,i} e^{-i E_n^{(i)} t} |\Phi_n^{(i)}\ra \la \Phi_n^{(i)} | \psi_{\rm in} \ra,
\label{psit}
\ee
where $E_n^{(i)}$ is the energy eigenvalue of the $\Phi_n^{(i)}$ state.

Using \eqref{Phins} and $\psi_{\rm in} = f_0$,
\be
\la \Phi_n^{(s)} | \psi_{\rm in} \ra = \frac{1}{\sqrt{2}} \delta_{n0} + \alpha \la g_n | \psi_{\rm in} \ra,
\ee
\be
\la \Phi_n^{(a)} | \psi_{\rm in} \ra = \frac{1}{\sqrt{2}} \delta_{n0} .
\ee
Eq.~\eqref{psit} gives the position space wavefunction
\ba
\psi (t,x) &=& 
\frac{e^{-i E_0^{(s)} t} }{\sqrt{2}} \biggl [ 
\la x |\Phi_0^{(s)} \ra + e^{-i \Delta E_0^{(a)} t} \la x |\Phi_0^{(a)}\ra
\nn \\
&& \hskip 0 cm
+ \alpha \sqrt{2} \sum_{n} e^{-i \Delta E_n^{(s)} t} \la g_n (x) | \psi_{\rm in} \ra \la x |\Phi_n^{(s)}\ra \biggr ] \\
&=&
\frac{e^{-i E_0^{(s)} t} }{\sqrt{2}} \biggl [ 
F_0^{\rm (s)} (x)  + e^{-i \Delta E_0^{(a)} t} F_0^{\rm (a)} (x)
\nn \\
&& \hskip -0.5 cm
+ \alpha \left \{ g_0(x) 
+ \sqrt{2} \sum_{n} e^{-i \Delta E_n^{(s)} t} \la g_n (x) | \psi_{\rm in} \ra F_n^{\rm (s)} (x) \right \} \nn \\
&&
+ {\cal O}(\alpha^2)
\biggr ],
\ea
where
\be
\Delta E_0^{(a)} \equiv E_0^{(a)} - E_0^{(s)}, \ \ 
\Delta E_n^{(s)} \equiv E_n^{(s)} - E_0^{(s)}.
\label{DeltaEdefns}
\ee

Noting the initial condition 
\be
\psi_{\rm in} = f_0 = (F_0^{\rm (s)} + F_0^{\rm (a)})/\sqrt{2},
\ee
we can write
\ba
\psi (t,x) &=& e^{-i E_0^{(s)} t}  \biggl [ 
f_0 (x) + \frac{(e^{-i \Delta E_0^{(a)} t} -1)}{\sqrt{2}} F_0^{\rm (a)} (x)
\nn \\
&& \hskip -1.9 cm
+ \alpha \sum_{n} (e^{-i \Delta E_n^{(s)} t}-1) \la g_n (x) | \psi_{\rm in} \ra F_n^{\rm (s)} (x)
+ {\cal O}(\alpha^2)
\biggr ].
\label{psit1}
\ea
A few features are apparent in this form. The first term in the square brackets in \eqref{psit1} is simply the initial condition. In the third term, the $n=0$ term does not contribute to the summation since $\Delta E_0^{(s)} =0$. The third term may be written as
\be
-2i\alpha \sum_{n} e^{-i \Delta E_n^{(s)} t/2} 
\sin (\Delta E_n^{(s)} t/2) \, g_n (x) F_n^{\rm (s)} (x) \nn,
\ee
so the summation only includes highly oscillating terms. The frequency of oscillation is set by $\Delta E_n^{\rm (s)}$ and is proportional to the spacing of the energy levels. These fast oscillations are in contrast to the second term with oscillation frequency set by $\Delta E_0^{\rm (a)} = {\cal O}(\alpha )$ which is parametrically small.

Ignoring the highly oscillating terms and expanding the exponential (since
$\Delta E_0^{\rm (a)} = {\cal O}(\alpha )$) we may write
\be
\psi (t,x) \sim e^{-i E_0^{(s)} t} \biggl [ 
f_0 (x) - \frac{i}{\sqrt{2}}  \Delta E_0^{(a)} F_0^{\rm (a)} (x) t
\biggr ].
\label{psitxexpanded}
\ee
From here we can evaluate the probability of the particle to be found in the $x >0$ region. Noting that $f_0(x)=0$ for $x > 0$, the probability for the particle to be found in the right-hand (``R'') well is
\ba
P_R (t) &=& \int_0^\infty dx \, |\psi (t,x)|^2 
\nn \\ &&
= \frac{(\Delta E_0^{(a)})^2 t^2}{2} \int_0^\infty dx \, ( F_0^{\rm (a)} (x) )^2 
\nn \\ &&
= \frac{1}{4} (\Delta E_0^{(a)})^2 t^2.
\label{PRt2}
\ea

The tunneling probability grows quadratically with time and only depends on 
\be
\Delta E_0^{(a)} = E_0^{(a)} -E_0^{(s)} = E_{\rm in} - E_0^{(s)}
\label{DeltaE0a}
\ee
This estimate only applies at early times as is clear from \eqref{psitxexpanded}. As time goes on and the particle moves into the right-hand well, there will be tunneling back to the left-hand well, setting up oscillations.

\section{Examples with symmetric potentials} \label{examplessymmetric}

\subsection{Infinite square well potential} \label{sqwell}

\begin{figure}
\includegraphics[width=0.30\textwidth,angle=0]{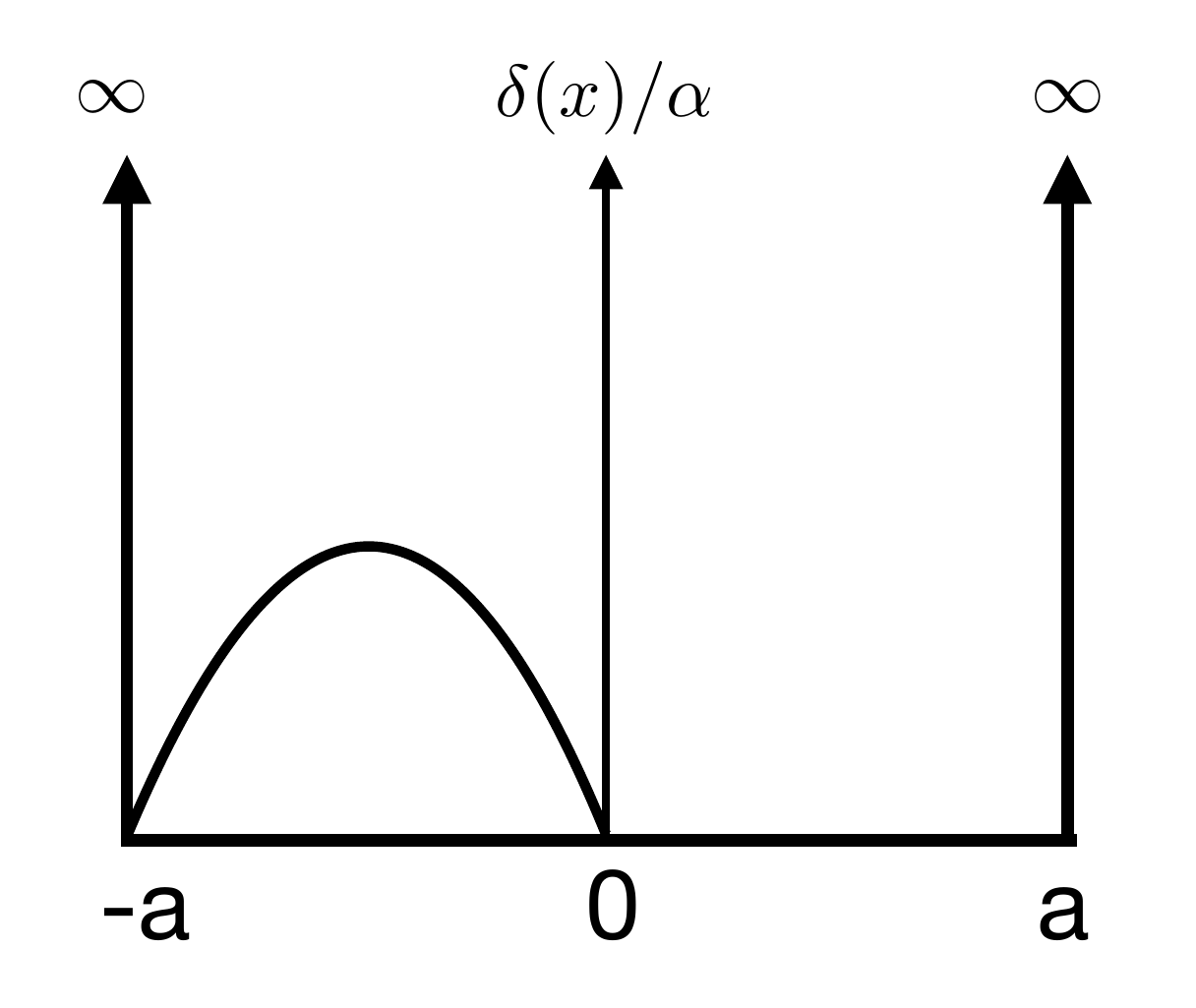}
\caption{The infinite square well with a delta function barrier at the center and a sketch of the initial wavefunction localized on the left-hand side of the barrier.}
\label{symsqwellfig}
\end{figure}

We now have the potential shown in Fig.~\ref{symsqwellfig}. The wavefunction for $\alpha =0$ is
\be
\psi_{\rm in}^{(j)}  = \sqrt{\frac{2}{a}} \, \sin(p (x+a)),
\ee
where $p = j \pi /a$ for the $j^{\rm th}$ mode. The energy (with $\alpha=0$) is
\be
E_{\rm in} = \frac{p^2}{2\mu} = \frac{j^2 \pi^2}{2\mu a^2}.
\ee
For small but non-zero $\alpha$ the wavefunction becomes
\be
\psi  = A \sqrt{\frac{2}{a}} \, \sin(k (x+a)),
\ee
where $A$ and $\delta$, which we still need to determine, are slight shifts in the normalization and the wavenumber due to $\alpha$ being non-zero (but still small). The energy is
\be
E_0^{(s)} = \frac{k^2}{2\mu} = \frac{1}{2\mu a^2}  (j \pi + \delta )^2
\approx E_{\rm in} + \frac{j\pi}{\mu a^2} \delta .
\label{E0ssqwell}
\ee

The boundary condition in \eqref{psi0} implies
\be
\tan (ka) =  - \alpha  \frac{k}{\mu}
\ee
which, to leading order in $\alpha$, gives
\be
\delta = -\alpha \frac{j \pi}{\mu a}.
\ee
From \eqref{DeltaE0a} and \eqref{E0ssqwell},
\be
\Delta E_0^{(a)} =  \alpha \frac{j^2 \pi^2}{\mu^2 a^3}
\ee
and from \eqref{PRt2},
\be
P_R^{(j)} (t) = \frac{\alpha^2}{4} \left ( \frac{ j\pi}{\mu a} \right )^4 \frac{t^2}{a^2}.
\label{sqwellPR}
\ee
The quantity in parentheses, $v_j \equiv j\pi/\mu a$, is the trapped particle's momentum divided by its mass, hence a velocity. Then $\tau_j =2\mu a^2/j\pi$ may be identified as the ``rattling time'', {\it i.e.} the time between successive encounters with the barrier. The formula can then be written as
\be
P_R^{(j)} (t) = \alpha^2 v_j^2 \left ( \frac{t}{\tau_j} \right )^2.
\label{sqwellPRinterpret}
\ee
It is tempting to identify $\alpha^2 v_j^2$ as the probability for the particle to escape with every hit, but this would lead to a linear growth of $P_R^{(j)} $ with $t$. The quadratic growth points to a resonance, which is natural given the symmetry of the problem and the perfect matching of energy levels in the left- and right- wells with $\alpha=0$. Indeed, we shall see in Sec.~\ref{asymV} that $P_R^{(j)}$ only grows linearly with $t$ when the energy levels do not match up.

\subsection{Quadratic potential} \label{quadpot}

We now take $V(x)=\mu \omega^2 x^2/2$ in \eqref{hamiltonian1}, and the initial wavefunction to be the first excited state of a simple harmonic oscillator for $x < 0$,
\be
\psi_{\rm in} (x) = \begin{cases}
\left ( \frac{\mu \omega}{\pi} \right )^{1/4} 2 \xi e^{-\xi^2/2} , \ \ x < 0 \\
0, \ \ x \ge 0 
\end{cases}
\label{psiin}
\ee
where $\xi = \sqrt{\mu \omega}\, x$ and we have taken care to normalize the wavefunction to unity. The energy corresponding to the initial state with $\alpha =0$ is
\be
E_{\rm in} = \frac{3}{2} \omega.
\ee

Next we consider the effect of non-zero but small $\alpha$ on the initial wavefunction.
We denote this wavefunction by $\Phi_0^{(s)}$ as in \eqref{Phins}, noting that
\be
F_0^{(s)}(x) = \frac{1}{\sqrt{2}} (\psi_{\rm in}(x) + \psi_{\rm in}(-x)).
\ee
The Schrodinger eigenvalue equation is
\be
-\frac{1}{2} \partial_\xi^2 \Phi_0^{(s)} + \frac{1}{2} \xi^2 \Phi_0^{(s)}
+ \frac{1}{\alpha} \sqrt{\frac{\mu}{\omega}} \, \delta (\xi ) \Phi_0^{(s)} = {\cal E} \Phi_0^{(s)}
\label{shoeq}
\ee
where ${\cal E}=E/\omega$.

We would like to solve \eqref{shoeq} with the boundary condition in \eqref{psi0} and with even parity: $\Phi_0^{(s)} (x) =   \Phi_0^{(s)}(-x)$. Equivalently we can solve \eqref{shoeq}
in $\xi \in [0,\infty )$ with the boundary condition
\be
\Phi_0^{(s)} (0) = \alpha \sqrt{\frac{\omega}{\mu}} \, \Phi_0^{(s)}{}' (0+),
\label{psibc}
\ee
where primes denote derivatives with respect to $\xi$. The solution for $\xi \in (-\infty,0]$ will follow from symmetry. The energy eigenvalue will be perturbed from ${\cal E} =3/2$, so we write
\be
{\cal E} = \frac{3}{2} + \alpha \epsilon
\ee
and we will work to first order in $\alpha$ to determine $\epsilon$.

From Appendix~\ref{shoE} we get,
\be
\Delta E_0^{(a)} = \frac{2\alpha }{\sqrt{\pi}}\sqrt{\frac{\omega}{\mu}} \, \omega
\ee
and, from \eqref{PRt2},
\be
P_R (t) = \frac{\alpha^2}{\pi} \frac{\omega^3}{\mu}  t^2 .
\label{shoPR}
\ee
This agrees with the square well estimate in \eqref{sqwellPR} up to numerical factors if we replace $a$ in \eqref{sqwellPR} by $1/\sqrt{\mu \omega}$ as this is the spread of the wavefunction in the quadratic potential.

\section{Tunneling in asymmetric potentials} \label{asymV}

\begin{figure}
\includegraphics[width=0.30\textwidth,angle=0]{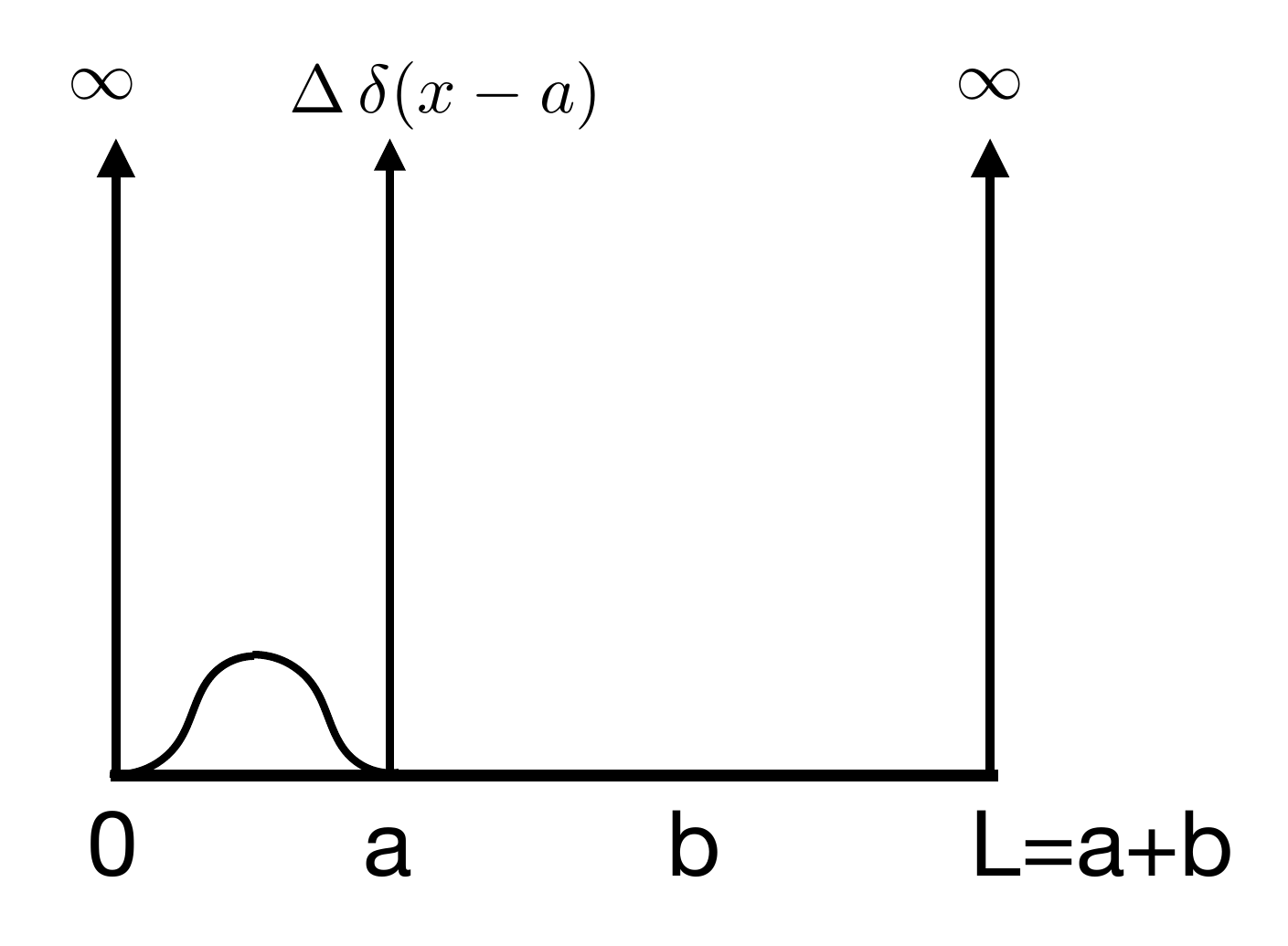}
\caption{Illustration of the asymmetric double square well potential.}
\label{potfig}
\end{figure}

The above sections dealt with double well situations in which the two wells are identical. We now consider asymmetric double wells but only the case of square wells (see Fig.~\ref{potfig}). The asymmetric case is quite different from the symmetric case as wavefunctions can no longer be split into even and odd parity modes.

The Hamiltonian for our setup is,
\be
{\hat H} = \frac{{\hat p}^2}{2\mu} + \alpha^{-1} \delta (x-a) + V_\infty(x)
\label{hamiltonianasym}
\ee
where 
\be
V_\infty (x) = \begin{cases}
\infty, & x \notin [0,L=a+b] \\
0. & x \in [0,L]
\end{cases}
\ee

We will solve the time-dependent Schrodinger equation  with initial conditions so that the particle is localized in the left-hand well. We choose the ratio $b/a$ of the widths of the two wells to be an irrational number. The irrationality simplifies the analysis because then the energy levels in the left- and right-hand
wells cannot be degenerate. 

Initially the particle is taken to be completely localized in the left well and we would like to calculate the probability of finding it in the right well as a function of time. We choose the wavefunction at the initial time ($t=0$) to be the $j^{\rm th}$ eigenstate of the left well
\be
\psi^{(j)} (t=0, x) = \begin{cases}
\sqrt{\frac{2}{a}} \sin \left ( j \frac{\pi}{a} x \right ), & x \in [0,a] \\
0. & x \in [a,L]
\end{cases}
\label{initial1}
\ee
The wavefunction outside of $[0,L]$ remains zero at all times. If one wishes, a general initial condition, still localized in the left hand well, can be constructed by  taking a linear combination of the eigenstates
\be
\psi (t=0,x) = \sum_j c_j \psi^{(j)} (t=0, x), 
\label{initialsuperposition}
\ee
where the coefficients $c_j$ can be chosen arbitrarily subject to the normalization condition: $\sum_j |c_j|^2 =1$.

\subsection{Irrational Double Well} \label{irrationalwells}

\subsubsection{Eigenstates} \label{eigenstatesirrational}

We first find the eigenstates of the full problem. These come in two sets: the first set is dominant in the left well, the second set is dominant in the right well. Irrationality of $b/a$ ensures that the energy levels of the left-hand and right-hand wells never match. This makes it easier to solve the matching conditions across the delta function barrier.

The eigenmodes in the first set are mostly in the left well and, to first order in $\alpha$, are
\ba
\Phi_{L,n} (t,x) &=& 
e^{-i E_{L,n} t}  \sqrt{\frac{2}{a}} \left (1+\frac{\delta_n}{2n\pi} \right )
\nn \\
&& \hskip -2 cm
\times
\begin{cases}
\sin(k_n x), & x \in [0,a]\\
(-1)^n \frac{\delta_n}{\sin (n\pi b/a )} \sin (k_n (L-x)), & x \in [a,L]
\end{cases}
\label{firstset}
\ea
where 
\ba
E_{L,n} &=& \frac{k_n^2}{2\mu}, \\
k_n &=& \frac{n\pi}{a} + \frac{\delta_n}{a}, \\
\delta_n &\approx& - \frac{n\pi \alpha}{2\mu a}.
\ea

The eigenmodes in the second set are mostly in the right well and, to first order in $\alpha$, are
\ba
\Phi_{R,n} (t,x) &=& 
e^{-i E_{R,n} t}  \sqrt{\frac{2}{b}}  \left (1+\frac{\epsilon_n}{2n\pi} \right )
\nn \\
&& \hskip -2 cm
\times
\begin{cases}
(-1)^n \frac{\epsilon_n}{\sin (n\pi a/b )} \sin (p_n x), & x \in [0,a] \\
\sin(p_n (L-x)), & x \in [a,L]
\end{cases}
\label{secondset}
\ea
where
\ba
E_{R,n} &=& \frac{p_n^2}{2\mu}, \\
p_n &=& \frac{n\pi}{b} + \frac{\epsilon_n}{b}, \\
\epsilon_n &\approx& - \frac{n\pi \alpha}{2\mu b}. 
\ea
Note that \eqref{firstset} and \eqref{secondset} contain some second order terms in $\alpha$ but these should be dropped and only first order terms should be retained.

\subsubsection{Evolution} \label{evolutionirrational}

We first write the wavefunction in terms of eigenmodes
\be
\Psi (t,x) = \sum_{n =1}^\infty a_n \Phi_{L,n} (t,x) + \sum_{n =1}^\infty b_n \Phi_{R,n} (t,x)
\label{expansion}
\ee
and find the coefficients $a_n^{(j)}$ and $b_n^{(j)}$ by using the initial conditions in
\eqref{initial1}:
\ba
a_n^{(j)} &=& \int_0^L dx\  \Phi_{L,n}^*(t=0,x) \psi (t=0,x) \\
&=&  \begin{cases}
1, & n=j \\
(-1)^{j+n} \frac{\delta_n}{\pi} \frac{2j }{n^2-j^2}, & n \ne j
\end{cases} 
\ea
\ba
b_n^{(j)} &=& \int_0^L dx\  \Phi_{R,n}^*(t=0,x) \psi (t=0,x) \\
&=& (-1)^{j+n} \sqrt{\frac{a}{b}} \, \frac{\epsilon_n}{\pi} \, \frac{2j}{ (n^2a^2/b^2-j^2)}.
\ea

The wavefunction in the right well, $x \in (a,L]$, is then,
\ba
\Psi_R^{(j)} (t,x) &=&  \alpha \frac{ (-1)^{j+1}  j \pi}{\sqrt{2} \mu a^{3/2} }  e^{-iE_{L,j} t}
\biggl [ \frac{\sin (j\pi (L-x)/a)}{\sin(j \pi b/a)}  \nn \\
&& \hskip -1.5 cm
- \frac{2}{\pi} 
\sum_{n=1}^\infty \frac{n \, e^{-i (E_{R,n}-E_{L,j} ) t}} {n^2 -j^2 b^2/a^2}
\sin (n\pi (x-a)/b ) \biggr ],
\label{psiR1}
\ea
where $E_{R,n}$ and $E_{L,j}$ are taken to zeroth order in $\alpha$. Using the initial condition that $\Psi_R^{(j)} (t=0,x)=0$ for $x \in (a,L)$ we can also write
\ba
\Psi_R^{(j)} (t,x) &=&  \alpha \frac{ (-1)^j \pi^2 \, j }{\sqrt{2} \mu^2 a^{3/2} b^2 } e^{-iE_{L,j} t} \nn \\
&& \hskip -2 cm
\times \sum_{n=1}^\infty \frac{n \, (e^{-i (E_{R,n}-E_{L,j} )t}-1)} {E_{R,n}-E_{L,j} }
\sin \left ( n\pi \frac{(x-a)}{b} \right ).
\label{psiR1again}
\ea
This formula applies only for $a < x \leq L$. In particular, the point $x=a$ is special since
\eqref{psiR1again} applied to $x=a$ gives zero, while \eqref{psiR1} gives
\be
\Psi_R^{(j)} (t=0,x=a) =  \frac{ (-1)^{j+1}  j \pi \alpha}{\sqrt{2} \mu a^{3/2} } \ne 0.
\label{PsiRjt0x}
\ee
This mismatch with the initial condition can be traced back to the first integral of the Schrodinger equation around $x=a$,
\be
-\frac{1}{2\mu} [\Psi'(t,a+)-\Psi'(t,a-)] + \alpha^{-1} \Psi(t,a) = i \int_{a-}^{a+}dx \, \partial_t \Psi.
\label{discontinuity}
\ee
Since there is a discontinuity in the first derivative of the initial wave function at $x=a$  and $\Psi (0,a)=0$, this relation implies that $\partial_t \Psi (t=0,a) \to \infty$. Thus there is a divergent transient at $t=0$ and the wavefunction at $t=0+$ need not agree with the wavefunction at $t=0$\footnote{This issue has nothing to do with the choice of a delta function barrier. It would occur even with a finite barrier but with an initial wavefunction that has a discontinuous first derivative.}. This issue can be mitigated by choosing an initial condition for which both the initial wavefunction and its first spatial derivative vanish at $x=a$. Such an initial condition, as discussed in Sec.~\ref{smoothinitial}, can be achieved by a suitable linear superposition of initial eigenmodes as in \eqref{initialsuperposition}, {\it i.e.} choice of coefficients $c_j$. However, even with smooth initial conditions at $x=a$, a transient cannot be avoided -- the wavefunction within the trap will evolve prior to tunneling 
taking place.

\begin{figure}
\includegraphics[width=0.45\textwidth,angle=0]{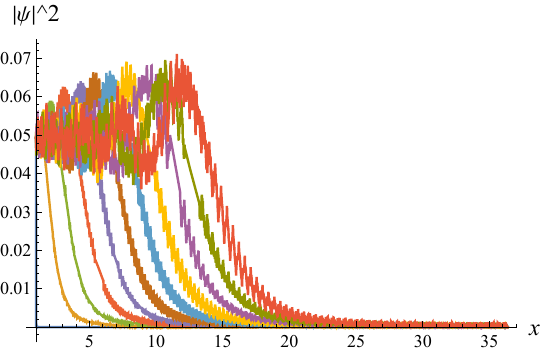}
\caption{Snapshots of $|\Psi_R^{(j)}|^2$ calculated from \eqref{psiR1} for $j=1$ with $\mu=1$, $a=1$, $b=25\sqrt{2}$, $\alpha =0.1$, $N=2000$, where $N$ is the upper cutoff on the sum over modes, $dt=0.5$ is the time step between plots, and we have taken 11 snapshots. The first snapshot at $t=0$ has $|\Psi_R^{(j)}(0,x)|^2 =0$ for $x >a$ and $|\Psi_R^{(j)}(0,a)|^2 \approx 0.05$. (Note that the origin is at $x=a=1$.)
}
\label{probfig}
\end{figure}

At $t=0$, the initial condition tells us that $\Psi_R^{(j)} (t,x)=0$ in the interval $x \in (a,L]$. As time goes on, $\Psi_R^{(j)}(t,x)$ becomes non-zero and the curve of $|\Psi_R^{(j)}(t,x)|^2$ lifts up and propagates into the right well as seen in a numerical evaluation of \eqref{psiR1} (see Fig.~\ref{probfig}). Within the present leading-order expression \eqref{psiR1}, the probability density at the $\delta$-function barrier is independent of time:
\be
|\Psi_R^{(j)} (t,x=a) |^2 =  \left ( \frac{ j \pi \alpha}{\sqrt{2}\, \mu a^{3/2} } \right )^2.
\ee

From \eqref{psiR1again} we can calculate the probability of the particle to be in the right-hand well as a function of time:
\ba
P_R^{(j)}(t) &=& \int_a^L dx \, |\Psi_R^{(j)} (t,x)  |^2 \nn \\
&& \hskip -1. cm
= \alpha^2 \frac{\pi^4 \, j^2 }{\mu^4 a^3 b^3 }
\sum_{n=1}^\infty n^2 \frac{\sin^2 ( (E_{R,n}-E_{L,j} )t/2)} {(E_{R,n}-E_{L,j})^2} \nn \\
&& \hskip -1 cm
= \alpha^2 \frac{\pi^2 \, j^2 }{\mu^3 a^3 b } t
\sum_{n=1}^\infty (\theta_n + E_{L,j}t/2) \frac{\sin^2 \theta_n} {\theta_n^2},
\label{PRjt}
\ea
where
\be
\theta_n \equiv
(E_{R,n}-E_{L,j} ) \frac{t}{2}.
\ee
By extremizing the summand as a function of $\theta_n$, assuming $\theta_n$ is a continuous variable and for large $E_{L,j}t/2$, we find a maximum at
\be
\theta_n = \theta_{j*} \sim \frac{3}{E_{L,j}t} \ll 1
\label{thetan}
\ee
corresponding to $n =n_{j*}$, where $n_{j*}$ denotes the integer that minimizes $( n^2-j^2b^2/a^2 )^2$. For large values of $jb/a$, 
\be
n_{j*}= [[(j b/a)+1/2]],
\ee
where $[[x]]$ denotes the greatest integer that is not greater than $x$. The formula for $\theta_{j*}$ can now be written in terms of $n_{j*}$, {\it i.e.} taking its discrete nature into account,
\ba
\theta_{j*} &=& \frac{\pi^2}{2\mu b^2} \left ( n_{j*}^2 -\frac{j^2 b^2}{a^2} \right ) \frac{t}{2}
\nn \\
&=&
 \frac{\pi^2}{2\mu b^2} \left ( \left [ \left [ \frac{jb}{a} +\frac{1}{2} \right ] \right]^2 
 -\frac{j^2 b^2}{a^2} \right ) \frac{t}{2}
\nn \\
&\sim&
 \frac{\pi^2}{2\mu b^2} \frac{j b}{2 a} \frac{t}{2} = \frac{\pi^2 j}{8\mu a b} t,
 \label{theta*}
\ea
where the last estimate is based on 
\be
\left | (j b/a) - [[(j b/a)+1/2]] \right | \sim 1/4
\label{floorest}
\ee
but the exact value depends on how close $j b/a$ is to being an integer.

A range of terms $\Delta n_j$ centered around $n_{j*}$ contribute to the sum in \eqref{PRjt}. $\Delta n_j$ is found by requiring that the argument of the sine function be order plus/minus
unity around the peak,
\be
\Delta n_j \sim \frac{8\mu a b}{j \pi^2 t} .
\label{Deltan}
\ee
Then, from \eqref{PRjt}, together with $E_{L,j} t \gg \theta_{j*}$, we find
\ba
P_R^{(j)}(t) &\sim& \frac{\alpha^2 2\pi^2 \, j^3 }{\mu^3 a^4}  \frac{\sin^2 \theta_{j*}} {\theta_{j*}^2} \, t,
\ \ t \gg E_{L,j}^{-1}
\nn \\
&=& \frac{2}{\pi} \alpha^2 v_j^2  \frac{\sin^2 \theta_{j*}} {\theta_{j*}^2} \frac{t}{\tau_L},
\label{PRirrational}
\ea
where $\tau_L = \mu a^2/\pi j$ is the rattling time in the left-hand well and $\theta_{j*}$ is $\theta_n$ at $n=n_{j*}$. Note that the expression in \eqref{PRjt} is even under $t \to -t$ while \eqref{PRirrational} is odd. The reason is that \eqref{Deltan} assumes $t > 0$; Eq.~\eqref{PRirrational} is really proportional to $t^2/|t|$.

Within the intermediate-time regime used to obtain \eqref{PRirrational} and while $\theta_{j,*}$ is small, $P_R^{(j)}(t)$ grows linearly with time. Eventually though $\theta_{j,*}$ gets large and $P_R^{(j)}(t)$ has oscillatory features. This occurs on a time scale given by the time it takes to rattle in the right-hand well, {\it i.e.} cross the  right-hand well and reflect back to the delta function barrier. This rattling time in the right-hand well is given by
\be
\tau_{R} = \frac{2b}{v_j} = \frac{2ab \mu}{j\pi}
\ee
since $v_j = j\pi/\mu a$ is the speed associated with the particle. From \eqref{theta*} we can write
\be
\theta_{j*} = \frac{\pi}{4} \frac{t}{\tau_R}.
\label{thetatauR}
\ee
Hence $\theta_{j*} \sim 1$ on a time scale of order $\tau_R$, when the particle can again tunnel back to the left well. However, the remark below \eqref{floorest} implies that there can be
numerical factors in \eqref{thetatauR} depending on how close $j b/a$ is to being an
integer.

We remark that in contrast to the result for the symmetric case given in \eqref{sqwellPR}, the probability of transfer to the right-hand well in the asymmetric case only grows linearly in $t$.

\subsubsection{Smooth initial state} \label{smoothinitial}

If we choose an initial state such that $\psi (0,a)=0$ and $\partial_x \psi (0,a) =0$, then \eqref{discontinuity} shows that the time derivative of the wavefunction at $x=a$ remains finite. A simple example of such an initial state is
\be
\psi (t=0, x) = \begin{cases}
\sqrt{\frac{8}{3a}} \sin^2 \left ( \frac{\pi}{a} x \right ), & x \in [0,a] \\
0. & x \in [a,L]
\end{cases}
\label{initialsmooth}
\ee
Then the coefficients $c_j$ in \eqref{initialsuperposition} can be evaluated,
\be
c_j = \frac{8}{\sqrt{3}\, \pi} \frac{ (-1)^j-1 }{j (j^2-4) },
\ee
and the probability for the particle to be found in the right-hand well is
\ba
P_R(t) &=& \sum_j |c_j|^2 P_R^{(j)} (t) \nn \\
&\sim& 
\frac{512 \alpha^2}{3\mu^3 a^4} \sum_{j={\rm odd}} \frac{j}{(j^2-4)^2}
\frac{\sin^2 \theta_{j*}} {\theta_{j*}^2} \, t .
\ea

\section{Escape to $\pm \infty$} \label{escape}

\begin{figure}
\includegraphics[width=0.50\textwidth,angle=0]{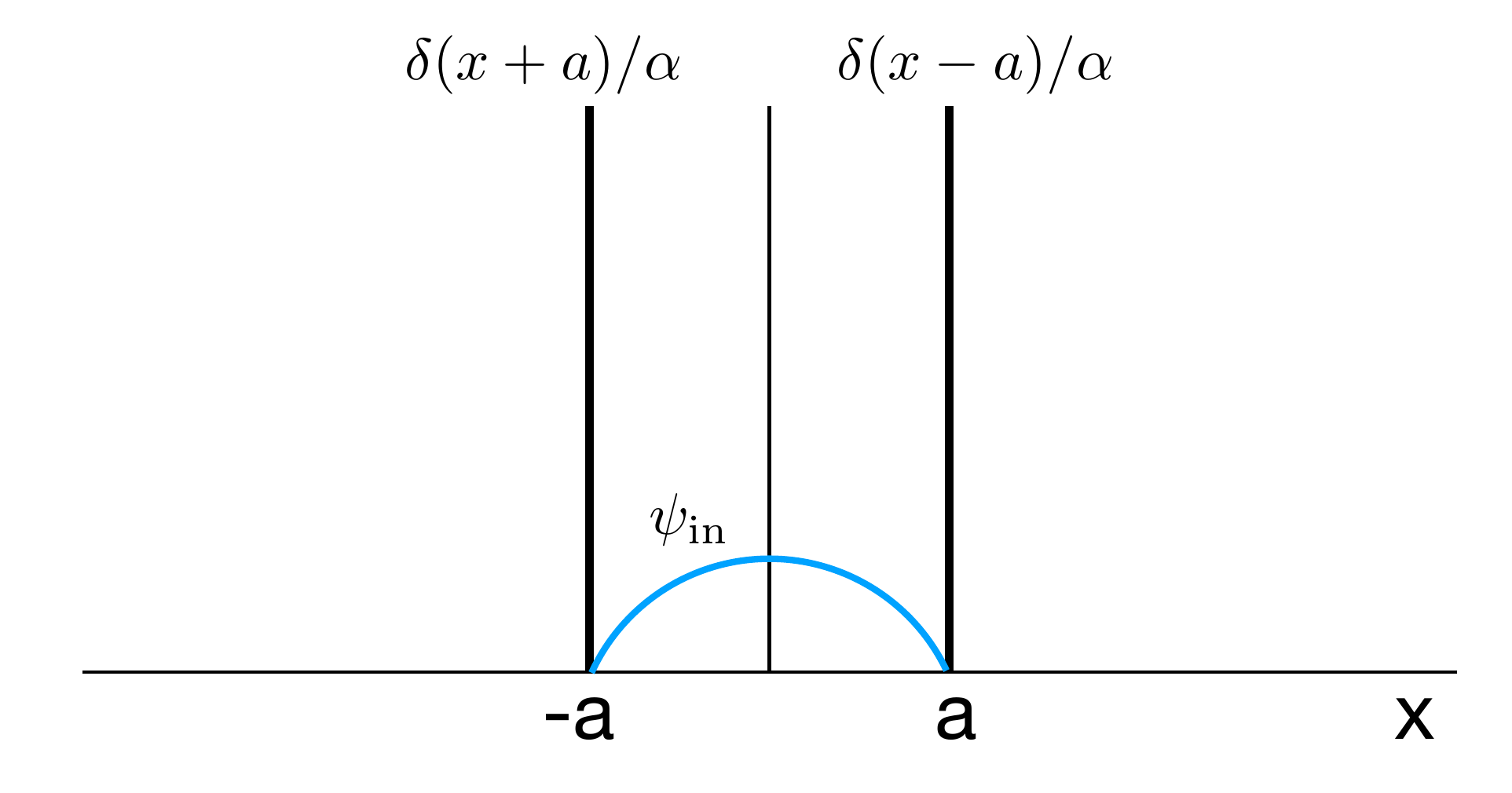}
\caption{The particle is initially trapped between two $\delta$-function barriers and eventually tunnels through to escape to $\pm \infty$.}
\label{escapefig}
\end{figure}

The Hamiltonian we consider is
\be
{\hat H} = \frac{p^2}{2\mu}  +  \alpha^{-1} \delta (x-a) +  \alpha^{-1} \delta (x+a)
\label{hamiltonian}
\ee
and the setup is illustrated in Fig.~\ref{escapefig}. The initial ($t=0$) wavefunction is chosen to be
\be
\psi_{\rm in} (x) = \frac{1}{\sqrt{a}}
\begin{cases}
\cos ( \pi x / (2a) ), & -a \le x \le a \\
0. & |x| > a
\end{cases}
\label{escapeinitial}
\ee

The energy eigenstates for the system are of the plane wave form with derivative discontinuities at the locations of the $\delta$-function barriers. Some algebra gives the parity even and parity odd eigenmodes,
\be
\Phi_k^{(e)} = \frac{1}{\sqrt{4\pi} \, |C_k|}
\begin{cases}
C_k e^{ikx} + C_k^* e^{-ikx}, & x < -a \\
e^{ikx} + e^{-ikx}, & -a \le x \le a \\
C_k^* e^{ikx} + C_k e^{-ikx}, & x > a
\end{cases}
\label{evenPhik}
\ee
\be
\Phi_k^{(o)} = \frac{1}{\sqrt{4\pi} \, |D_k|}
\begin{cases}
D_k e^{ikx} - D_k^* e^{-ikx}, & x < -a \\
e^{ikx} - e^{-ikx}, & -a \le x \le a \\
D_k^* e^{ikx} - D_k e^{-ikx}, & x > a
\end{cases}
\label{oddPhik}
\ee
where 
\ba
C_k &=& 1+i \beta_k (1+ e^{i2ka}), \label{Ck} \\
D_k &=& 1+i \beta_k (1-e^{i2ka}), \label{Dk}
\ea
with $\beta_k \equiv \mu/(\alpha k)$. The energy eigenmodes satisfy $\delta$-function normalization, for example,
\ba
\int dx \, \Phi_k^{(e)*}(x) \Phi_p ^{(e)}(x)  &=& \delta(k-p) \nn \\
&=& \int dx \, \Phi_k^{(o)*}(x) \Phi_p ^{(o)} (x),
\label{orthonormal}
\ea
while the overlap of even modes with odd modes vanishes because of parity.

The energy eigenvalues are as for a free particle,
\be
E_k = \frac{k^2}{2\mu}.
\ee
We will only consider $k \ge 0$ to avoid double counting eigenmodes.

Now we express the initial wavefunction in terms of the eigenmodes (at $t=0$),
\be
\psi_{\rm in} (x) = \int_0^\infty dk \, [ c_k \Phi_k^{(e)} + d_k \Phi_k^{(o)} ].
\label{psiinPhi}
\ee
Using orthonormality of the eigenmodes, Eq.~\eqref{orthonormal},
\ba
c_k &=& \int dx \, \Phi_k^{(e)*} (x) \psi_{\rm in} (x)  \nn \\
&=& \frac{\sqrt{\pi a}}{ |C_k|} \frac{\cos(ka)}{(\pi/2)^2- (ka)^2}.
\label{ckcoeffs}
\ea
We also note
\be
|C_k|^2 = 1 - 2 \beta_k \sin(2ka) +4\beta_k^2 \cos^2(ka) .
\label{Ck2}
\ee
The coefficient $d_k=0$ because there is no overlap of $\psi(0,x)$ with the parity odd eigenmodes.

Completeness of the mode functions in \eqref{evenPhik} and \eqref{oddPhik} implies
\be
\int_0^\infty dk [ \Phi_k^{(e)*}(x) \Phi_k^{(e)} (y) + \Phi_k^{(o)*}(x) \Phi_k^{(o)} (y) ] = \delta(x-y).
\ee
We have not been able to show this relation explicitly. However, we have numerically checked  that \eqref{psiinPhi} is satisfied with $c_k$ given in \eqref{ckcoeffs} and $d_k=0$, which is all that we need.

\subsection{Evolution inside the trap} \label{evolutioninside}

The time evolved wave function inside the trap, {\it i.e.} the $|x| < a$ region, is now given by
\ba
\psi(t,|x| < a ) &=& \int_0^\infty dk \,  e^{-iE_k t} c_k \Phi_k^{(e)} (|x|<a) \nn \\
&\equiv& K(t,x) + K^*(-t,x),
\label{psiinside}
\ea
where
\be
K(t,x) =
\frac{1}{2\sqrt{a}} \int_0^{+\infty} d\kappa \, e^{-iE_\kappa \tau} e^{i\kappa X} 
\frac{1}{|C_k|^2} \frac{\cos(\kappa)}{(\pi/2)^2- \kappa^2}
\label{Ktx}
\ee
and we have defined dimensionless variables $\kappa =ka$, $X=x/a$, $E_\kappa = \kappa^2/2M$,  $M=\mu a$ and $\tau =t/a$. 

Let us consider the function
\ba
f(\kappa) &=&  \alpha^2 \kappa^2 |C_k|^2 \nn \\
&=& \alpha^2 \kappa^2 - 2M \alpha \kappa \sin(2\kappa) + 4 M^2 \cos^2(\kappa) .
\label{fkappadefn}
\ea
As discussed in Appendix~\ref{estimatingintegral}, the function $f(\kappa)$ has a discrete set of sharp minima for small $\alpha$. The maximum contribution to the integral in \eqref{Ktx} in the limit of small $\alpha$ comes from the region around the location of the minimum of $f(\kappa)$ (see \eqref{kappa*}),
\be
\kappa_* = \frac{\pi}{2} \left ( 1 - \frac{\alpha}{2M} + \frac{\alpha^2}{4M^2} \right ) + {\cal O}(\alpha^3).
\label{kappa*appB}
\ee
Around the minimum we can take,
\ba
f(\kappa) &=& \frac{\pi^4 \alpha^4}{64 M^2} + 4 M^2 u^2 + {\cal O}(u^3) \nn \\
&=& 4 M^2 (\sigma^2 + u^2) + {\cal O}(u^3),
\ea
where $u = \kappa - \kappa_*$
and
\be
\sigma \equiv \left ( \frac{\alpha \pi}{4M} \right )^2.
\label{sigmadefnappB}
\ee
Then we approximate
\be
\frac{1}{|C_k|^2} \frac{\cos(\kappa)}{(\pi/2)^2- \kappa^2} \approx 
\frac{\alpha^2 \pi}{16 M^2} \frac{1}{\sigma^2 + u^2},
\ee
\ba
E_k t -kx &=& \frac{k^2}{2\mu} t - k x  \nn \\
&& \hskip -2.0 cm
= \left ( \frac{\kappa_*^2 \tau}{2M} - \kappa_* X \right ) + (v_* \tau - X) u + {\cal O}(u^2),
\label{Ekt-kx}
\ea
where $v_* = \kappa_*/M = \pi/ 2\mu a$ denotes the speed of the particle. This is reasonable because the typical momentum of the particle within the trap is $\approx \kappa_*$ and so the velocity is $v_* = \kappa_* /M$. We can drop the ${\cal O}(u^2)$ terms in \eqref{Ekt-kx} as long as
\be
\frac{2M}{\tau} |v_* \tau - X| \gg |u|.
\label{uapprox}
\ee
With these approximations we obtain
\ba
K(t,x) &\approx&
\frac{\alpha^2 \pi e^{-i(E_* \tau -\kappa_* X)} }{32 M^2\sqrt{a}} 
\int_{-\infty}^{+\infty} du \, \frac{e^{-i(v_*\tau - X)u}}{u^2+\sigma^2} \nn \\
&=&
\frac{1}{2\sqrt{a}} e^{-i(E_* \tau -\kappa_* X)} e^{- \sigma |v_* \tau - X|}
\label{Ktxapprox}
\ea
with $E_* = \kappa_*^2/2M$. For $|v_* \tau - X | < 1/\sigma$ the integral gets contributions from  $|u| \sim \sigma$, in which case \eqref{uapprox} shows that the approximation holds for $|X - v_* \tau | \gg (\sigma/2M)\tau$.

Eq.~\eqref{psiinside} now gives the wavefunction inside the trap,
\ba
\psi(t,|x| < a ) &\approx& \frac{1}{2\sqrt{a}} e^{-iE_*\tau} \biggl [
e^{-\sigma |v_*\tau -X|} e^{i\kappa_* X} \nn \\
&& \hskip 1 cm
+ e^{-\sigma |v_*\tau +X|} e^{-i\kappa_* X} \biggr ].
\label{psiinsideresult}
\ea

We can check this solution explicitly by inserting \eqref{psiinsideresult} into the Schrodinger equation with the Hamiltonian in \eqref{hamiltonian}. We find
\be
H\psi - i \partial_t \psi = \frac{\sigma e^{+iE_* \tau}}{2Ma^{3/2}} 
\left [ \delta(X-v_* \tau) + \delta(X + v_* \tau) \right ] + {\cal O}(\sigma^2)
\ee
where $-1 < X < +1$. The right-hand side vanishes inside the trap (to order $\sigma^2$) for $\tau > 1/v_*$. This verifies that our solution in \eqref{psiinsideresult} solves the Schrodinger equation and shows that it is valid for
\be
t > \frac{\mu a^2}{\kappa_*}.
\ee

We can now evaluate the probability of the particle to remain inside the trap,
\ba
P_{\rm in}(t) &=& \int_{-a}^{+a} dx\, |\psi (t, |x|<a)|^2 \nn \\
&=&
\frac{\sinh(2\sigma) }{2\sigma} e^{-2\sigma v_*\tau} \approx e^{-2\sigma v_*\tau}.
\ea
Therefore the escape probability is
\be
P_{\rm esc} (t) \approx 1-e^{-2\sigma v_*\tau} 
\label{Pesc}
\ee
and the decay time is
\be
t_{\rm decay} = \frac{a}{2\sigma v_*} = \frac{16 \mu^3 a^4}{\alpha^2 \pi^3} = \frac{\tau_*}{\alpha^2 v_*^2},
\label{tdecay}
\ee
where $\tau_* = 2a/v_*$ is the rattling time within the trap. In line with the discussion below
\eqref{sqwellPRinterpret}, we can think of the particle rattling inside the trap with an escape
probability given by $\alpha^2 v_*^2$ per hit on the trap walls.

\subsection{Wavefunction outside the trap}
\label{evolutionoutside}

The time evolved wave function in the $x > a$ region is given by
\ba
\psi(t,x >a ) &=& \int_0^\infty dk \,  e^{-iE_k t} c_k \Phi_k^{(e)} (x>a) \nn \\
&& \hskip -2.5 cm
= \frac{\sqrt{a}}{2} \int_0^{+\infty} dk\, e^{-iE_k t} e^{ikx} \frac{C_k^*}{|C_k|^2}
\frac{\cos(ka)}{(\pi/2)^2- (ka)^2} \nn \\
&& \hskip -2.5 cm
+ \frac{\sqrt{a}}{2} \int_0^{+\infty} dk\, e^{-iE_k t} e^{-ikx} \frac{C_k}{|C_k|^2}
\frac{\cos(ka)}{(\pi/2)^2- (ka)^2} \nn \\
&& \hskip -2.5 cm
\equiv I(t,x) + I^* (-t,x).
\label{psitxa1}
\ea

We now use \eqref{Ck} to write,
\be
C_k^* = \frac{1}{\alpha \kappa} [ (\alpha \kappa - M \sin(2\kappa)) - i M (1+\cos(2\kappa))]
\ee
and, from Appendix~\ref{estimatingintegral},
\be
|C_k|^2 =\frac{1}{\alpha^2 \kappa^2} f(\kappa) 
\simeq \frac{4M^2}{\alpha^2\kappa^2} (\sigma^2 + u^2)
\ee
where $u=\kappa-\kappa_*$.

Next we expand terms in the integrand in powers of $\alpha$,
\ba
&&
\alpha \kappa - M \sin(2\kappa) \approx -M\sin(2u), \nn \\
&&
1+\cos(2\kappa) \approx 2 \sin^2 u , \nn \\
&&
\cos\kappa \approx -\sin u , \nn \\
&&
(\pi /2)^2 - \kappa^2 \approx -\pi u ,
\ea
and we have assumed that $|u| \ll 1$.
With these expansions we find
\be
C_k^* \approx \frac{-2M }{\alpha \kappa_*} \sin u \, e^{i u} .
\ee
In addition,
\ba
E_k t -kx &=& \frac{k^2}{2\mu} t - k x  \nn \\
&& \hskip -2.0 cm
= \left ( E_* \tau - \kappa_* X \right ) + (v_* \tau - X) u + {\cal O}(u^2)
\ea
where $\tau \equiv t/a$, $X\equiv x/a$, $E_*\equiv \kappa_*^2/2M$ and $v_* = \kappa_*/M$.

Putting these approximations together we get
\ba
I(t,x>a) &\approx& i \frac{\alpha e^{-i(E_* \tau -\kappa_* X)}}{16 M \sqrt{a}} 
\int^{\infty}_{-\infty} \frac{du}{u^2 + \sigma^2} 
 \nn \\
&& \hskip -1.5 cm
\times \frac{\sin u}{u} \left [ e^{-i (v_* \tau - X -2) u} - e^{-i (v_*\tau -X)u} \right ]
\ea

The integral gets contributions from one of the two poles at $u=\pm i \sigma$
leading to,
\ba
I(t,x>a) &\approx& 
i \frac{\pi \alpha}{16 M \sqrt{a}} e^{-i(E_* \tau -\kappa_* X)} \nn \\
&& \hskip -0.5 cm
\times \frac{1}{\sigma} \left [ e^{-\sigma | v_* \tau - X -2 |} - e^{-\sigma | v_*\tau -X |} \right ]
\ea
from which the wavefunction follows using \eqref{psitxa1}.

\section{Relation to Other Approaches} \label{other_approaches}

The calculations above are closely related to several established approaches to time-dependent tunneling. Delta-function barriers, resonant tunneling, metastable decay, and tunneling transients have all been studied extensively using exact and systematic methods~\cite{Kleber_1994, Del_Campo_2009}. Our aim here is to isolate a simple asymptotic regime in which the barrier is tall and thin and its integrated strength provides a natural control parameter. In this limit, coherent transfer between nearly degenerate bound states, nonresonant leakage, and escape into a continuum can be treated and compared in a unified way.

\subsection{Green functions, poles, and resonances}

A standard way to analyze time-dependent tunneling is through the retarded Green function or propagator. In that language the time-dependent wavefunction is represented schematically as
\begin{equation}
\psi(x,t)=\int dE\, e^{-iEt}\,G(x,x';E)\,\psi(x',0),
\end{equation}
with the relevant physical regimes encoded in the analytic structure of $G(E)$. Bound states appear as poles on the real axis, resonances or quasi-bound states as poles continued into the lower half of the complex energy plane, and continuum effects through branch cuts. Deforming the energy contour then separates the time evolution into pole contributions and background-continuum contributions.

This point of view gives a useful interpretation of the results above. In the symmetric double-well examples the dominant dynamics occur within an almost degenerate two-dimensional subspace. The tunneling is therefore governed by the splitting between the even and odd states, and the transition probability grows quadratically at early times before becoming part of a coherent oscillation. In the asymmetric case the absence of exact level matching suppresses this resonant two-state reduction, and the transfer resembles weak nonresonant leakage into a set of detuned states. In the escape problem, where a state initially localized in the trap couples to a continuum, the usual pole contribution gives exponential decay over its regime of validity, while the non-pole background controls the early-time transient and late-time corrections. Thus the Green-function perspective organizes the calculation in terms of discrete eigenvalues, resonance poles, and branch cuts. 


\subsection{Moshinsky functions and diffraction in time}

There is also a close connection with the literature on quantum transients and diffraction in time, initiated by Moshinsky's shutter problem~\cite{Moshinsky_1952}. In the simplest shutter setup, a wave initially cut off by a shutter is suddenly released, and the subsequent time-dependent solution is expressed in terms of the Moshinsky function,
\begin{equation}
M(x,k,t) = \frac{1}{2} e^{ikx-i k^2 t/(2\mu)} \operatorname{erfc}
\left[
\frac{x-k t/\mu}{\sqrt{2 i t/\mu}}
\right],
\end{equation}
up to conventional normalizations. This function is the natural time-domain kernel for a sharply released nonrelativistic wave. It describes the transient buildup of amplitude, including the oscillatory ``diffraction in time'' structure near the propagating front.

Delta-function barriers are especially close to this framework. For a barrier
\begin{equation}
V(x)=\lambda \delta(x),
\end{equation}
the stationary transmission amplitude has a simple rational dependence on momentum. When the time-dependent transmitted wave is reconstructed by Fourier synthesis, the resulting integrals can be expressed in terms of Moshinsky functions, or equivalently in terms of pole contributions dressed by the causal transient kernel. In this sense the thin-barrier limit used here naturally intersects the Moshinsky-function formulation of time-dependent scattering.

The connection is most direct for tunneling into extended states or escape into the continuum, where the outgoing wave can be written as an integral over continuum momenta. The Moshinsky representation then makes explicit how the transmitted amplitude is built up in time. For bound-to-bound tunneling, however, the more economical description is a projection onto a nearly degenerate discrete subspace; there the same thin barrier appears through the induced level splitting, rather than through a continuum shutter kernel. Thus the Moshinsky formalism and the two-state tunneling description emphasize different projections of the same underlying dynamics.

This distinction also clarifies the different time dependences encountered above. Strictly at very short times, transition probabilities in quantum mechanics are generically quadratic when the relevant energy variance is finite. Linear-in-time behavior is better understood as a coarse-grained or intermediate weak-coupling regime associated with leakage into many nonresonant states, while exponential decay arises when a resonance pole dominates the continuum evolution. The thin-barrier model provides a simple setting in which these different regimes can be displayed side by side.

Our contribution, therefore, is not the introduction of new special functions or a new analytic machinery for time-dependent tunneling. Rather, it is to use the tall, thin barrier as a controlled and physically transparent limit in which the familiar Green function and Moshinsky structures reduce to elementary formulae for resonant transfer, nonresonant leakage, and escape.

\section{Conclusions} \label{conclusions}

We have developed a perturbative technique to analyze tunneling through a tall, thin barrier. The parameter that controls the tunneling rate is denoted $\alpha$ and is given by the inverse of the potential function integrated between the turning points (see Eq.~\eqref{alpha-1}). There is no tunneling when $\alpha =0$ and weak tunneling for small $\alpha$. This is a different regime from the usual treatment of tunneling in the WKB approximation when the tunneling {\it action} is taken to be large. In fact, small values of the barrier width for fixed $\alpha$ imply small tunneling action (see Eq.~\eqref{kappad}).

Using the general perturbative treatment around a delta function barrier in Sec.~\ref{general}, we explicitly solved for tunneling in a potential with a tall, thin barrier placed so as to create a {\it symmetric} double well configuration. The results for the examples of the square well and the quadratic potentials in Sec.~\ref{examplessymmetric} lead to the same parametric result  that the tunneling probability from one well to the other grows as $t^2$ at early times as in \eqref{shoPR}. The quadratic behavior is due to resonances between the degenerate energy levels on the two sides of the barrier. The expression for the probability of the particle to be found in the right hand well, 
Eq.~\eqref{sqwellPRinterpret}, is
\be
P_R^{(j)} (t) = \alpha^2 v_j^2 \left ( \frac{t}{\tau_j} \right )^2,
\ee
where $v_j$ is the velocity of the particle within the well and $\tau_j = 2a/v_j$ is the rattling time within the trap.

In contrast, in the case of {\it asymmetric} wells with irrational width ratio, discussed in Sec.~\ref{asymV}, resonances are absent and the tunneling probability grows linearly with time, as in
\eqref{PRirrational},
\be
P_R^{(j)}(t) \sim \frac{2\alpha^2 v_j^2}{\pi} \, \frac{\sin^2 \theta_{j*}} {\theta_{j*}^2} \,  \frac{t}{\tau_L},
\ \ t \gg E_{L,j}^{-1}
\ee
where the angle $\theta_{j*}$ depends on the mismatch between the energy level within the trap and the closest energy level outside the trap (see the discussion above \eqref{floorest}). For $\theta_* \ll 1$, the characteristic transfer time in the asymmetric case is
\be
t_{\rm asym} = \frac{\pi \tau_L}{2 \alpha^2 v_j^2},
\label{tassel}
\ee
where $\tau_L$ is the rattling time in the trapping (left-hand) well.

Our last example was of tunneling out from a trap and escape to infinity. Now there is a set of bound states within the trap and a continuum of states outside the trap, with weak interactions between the two due to the delta function barrier. In this case, the escape probability is given by the usual exponential formula (see \eqref{Pesc}) and the escape time scale is given by \eqref{tdecay} which can also be written as
\be
t_{\rm decay} = \frac{\tau_*}{\alpha^2 v_*^2}
\label{tdecayagain},
\ee
where $\tau_* = 2a/v_*$ is the rattling time and $v_* = \pi/2\mu a$ is the velocity of the particle. This expression is very similar to the time scale we would associate to the asymmetric case but without any factor due to the mismatch of energy levels. Besides the escape time, we have explicitly evaluated the wavefunction of the tunneling particle inside (see Eq.~\eqref{psiinsideresult}) and outside the trap (Sec.~\ref{evolutionoutside}).

The idealized models considered here may be useful as benchmarks for engineered systems in which narrow tunable barriers are natural, such as ultracold atom systems with optically generated barriers~\cite{Albiez_2005, Jendrzejewski_2016, Eid_2024}, gate-defined mesoscopic structures~\cite{Van_Der_Wiel_2002}, or simplified models of leakage from traps. In such contexts one often wants transparent estimates of transient transfer or escape without performing a full numerical time evolution. 

Let us also mention that generalized kinetic terms beyond the usual quadratic dispersion can lead to significant physical effects~\cite{Wilczek_2025}. Rapid changes in momentum, such as occur at sharp barriers, can enhance the effects of these terms. Such terms introduce calculable corrections to the formulas derived above.

\acknowledgements
TV thanks Nicola Garofalo for discussions and
the Scirce project’s review agent for a careful reading of the manuscript and feedback.
This work was supported by the U.S. Department of Energy, Office of High Energy 
Physics, under Award No.~DE-SC0019470.

\appendix

\section{SHO energy correction}
\label{shoE}

Here we evaluate the energy of $\Phi_0^{(s)}$ as in \eqref{Phins} to lowest order in $\alpha$. As in \eqref{Phins} we write,
\be
\Phi_0^{\rm (s)} (x) = F_0^{\rm (s)}(x) + \alpha g_0(x)
\ee
The energy eigenvalue is written as
\be
{\cal E} = \frac{3}{2}+ \alpha \epsilon .
\ee
Then the equations satisfied by $F_0^{(s)}$ and $g_0$ are,
\ba
\partial_\xi^2 F_0^{(s)} - \xi^2 F_0^{(s)} &=& -3 F_0^{(s)} \nn \\
\partial_\xi^2 g_0 - \xi^2 g_0 &=& -3 g_0 - 2 \epsilon F_0^{(2)}
\ea
We evaluate the Wronskian and integrate over $\xi \in (0,\infty)$ to obtain,
\be
(g_0 \partial_\xi F_0^{(s)} - F_0^{(s)} \partial_\xi g_0 ) \big |_0^\infty = 2\epsilon \int_0^\infty |F_0^{(s)}|^2 d\xi
\label{wrons}
\ee
Since $F_0^{(s)}$ is normalized to unity over $x \in (-\infty, +\infty)$ and $\xi = \sqrt{\mu \omega} \, x$, the integral on the right-hand side is $\sqrt{\mu\omega}/2$. Additionally we know that
\be
F_0^{(s)} (0)=0=F_0^{(s)}(\infty) = \partial_\xi F_0^{(s)} (\infty)
\ee
and the boundary condition in \eqref{psibc} gives
\be
g_0(0) = \sqrt{\frac{\omega}{\mu}} \, F_0^{(s)}(0+).
\ee
Inserting these values into \eqref{wrons} and \eqref{Fns} gives
\be
\epsilon = - \frac{1}{2\mu} | \partial_\xi \psi_{\rm in}(0-)|^2 = - \frac{2}{\sqrt{\pi}} \sqrt{\frac{\omega}{\mu}}
\ee
where $\psi_{\rm in}$ is defined in \eqref{psiin}. Therefore
\be
\Delta E_0^{(a)} = - \alpha \epsilon \omega = \frac{2\alpha }{\sqrt{\pi}}\sqrt{\frac{\omega}{\mu}} \, \omega .
\ee

\section{Estimating $f(\kappa)$ of Sec.~\ref{escape}} \label{estimatingintegral}

Equation~\eqref{fkappadefn} defines
\be
f(\kappa) = \alpha^2 \kappa^2 - 2M \alpha \kappa \sin(2\kappa) + 4 M^2 \cos^2(\kappa) .
\ee
Taking derivatives with respect to $\kappa$ we get
\be
f'(\kappa) = 2\alpha^2\kappa - 2M (\alpha+2M) \sin(2\kappa) - 4M\alpha \kappa \cos(2\kappa),
\ee
\be
f''(\kappa) = 2\alpha^2 -8M(\alpha+M)\cos(2\kappa) + 8M\alpha \kappa \sin(2\kappa).
\ee

\begin{figure}
\includegraphics[width=0.40\textwidth,angle=0]{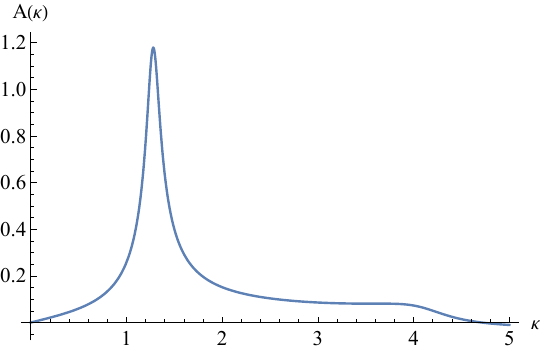}
\caption{A plot of the function ${\cal A}(\kappa)$ in \eqref{Akappa} vs. $\kappa$ for $M=1$, $\alpha=0.5$.}
\label{Akvsk}
\end{figure}

For $\alpha=0$, $f(\kappa)$ has extrema at $\sin(2\kappa)=0$, {\it i.e.} at $\kappa = n\pi/2$ where $n$ is an integer. Because of the other factors in \eqref{Ktx} and \eqref{Akappa},  the integrands are dominated by the $n=\pm 1$ minima (see Fig.~\ref{Akvsk}) and we will focus on the $n=+1$ case (the $n=-1$ expressions can be obtained by letting $\kappa \to -\kappa$). For small but non-zero $\alpha$, the positions of the $n=+1$ minima will shift and will occur at
\be
\kappa_* = \frac{\pi}{2} + z
\ee
where $z= {\cal O}(\alpha )$. Working to quadratic order in $\alpha$, $f'(\kappa_*)=0$ gives
\be
z = - \frac{\pi}{2} \left ( \frac{\alpha}{2M} - \frac{\alpha^2}{4M^2} \right ) + {\cal O}(\alpha^3)
\ee
and hence,
\be
\kappa_* \approx \frac{\pi}{2} \left ( 1 - \frac{\alpha}{2M} + \frac{\alpha^2}{4M^2} \right ) + {\cal O}(\alpha^3).
\label{kappa*}
\ee
Again, working to lowest order in $\alpha$, we find
\be
f(\kappa_*) = \frac{\pi^4 \alpha^4}{64 M^2}  + {\cal O}(\alpha^5)
\ee
and
\be
f''(\kappa_*) = 8M^2 + {\cal O}(\alpha).
\ee

Denoting $u = \kappa-\kappa_*$, the
Taylor expansion of $f(\kappa)$ around $\kappa =\kappa_*$ gives
\ba
f(\kappa) &=& \frac{\pi^4 \alpha^4}{64 M^2} + 4 M^2 u^2 + {\cal O}(u^3) \nn \\
&=& 4 M^2 (\sigma^2 + u^2) + {\cal O}(u^3) 
\label{approxfkappa}
\ea
where
\be
\sigma \equiv \left ( \frac{\alpha \pi}{4M} \right )^2.
\label{sigmadefn}
\ee

\bibstyle{aps}
\bibliography{paper}

\end{document}